\documentclass[twocolumn,showpacs,pre,superscriptaddress]{revtex4}

\usepackage{graphicx}
\usepackage[usenames]{color}
\usepackage{amssymb}
\usepackage{amsmath}
\usepackage{amsfonts}
\usepackage{mathrsfs}
\renewcommand{\epsilon}{\varepsilon}
\newcommand{\figurewidth}{0.46\textwidth}

\begin{document}
\title{Polymer translocation out of confined environments}

\author{Kaifu Luo}
\altaffiliation[]{ Author to whom the correspondence should be addressed}
\email{luokaifu@gmail.com}
\affiliation{Physics Department, Technical
University of Munich, D-85748 Garching, Germany}
\author{Ralf Metzler}
\affiliation{Physics Department, Technical University of Munich,
D-85748 Garching, Germany}
\author{Tapio Ala-Nissila}
\affiliation{Department of Applied Physics, Helsinki University of Technology,
P.O. Box 1100, FIN-02015 TKK, Espoo, Finland} \affiliation{Department of
Physics, Box 1843, Brown University, Providence, Rhode Island 02912-1843, USA}
\author{See-Chen Ying}
\affiliation{Department of Physics, Box 1843, Brown University, Providence,
Rhode Island 02912-1843, USA}

\date{July 7, 2009}

\begin{abstract}

We consider the dynamics of polymer translocation out of confined environments.
Analytic scaling arguments lead to the prediction that the translocation time
scales like $\tau\sim N^{\beta+\nu_{2D}}R^{1+(1-\nu_{2D})/\nu}$ for
translocation out of a planar confinement between two walls with separation $R$
into a 3D environment, and $\tau \sim N^{\beta+1}R$ for translocation out of
two strips with separation $R$ into a 2D environment. Here, $N$ is the chain
length, $\nu$ and $\nu_{2D}$ are the Flory exponents in 3D and 2D, and $\beta$
is the scaling exponent of translocation velocity with $N$, whose value for the
present choice of parameters is $\beta \approx 0.8$ based on Langevin dynamics
simulations. These scaling exponents improve on earlier predictions.


\end{abstract}


\maketitle
\section{Introduction}
%
The transport of a polymer through a nanopore has received increasing attention
due to its importance in biological systems~\cite{Alberts}, such as  gene
swapping through bacterial pili, m-RNA transport through nuclear pore
complexes, and injection of DNA from a virus head into the host cell, etc.
Moreover, translocation processes have various potentially revolutionary
technological applications~\cite{Meller03,Kasianowicz}, such as rapid DNA
sequencing, gene therapy, and controlled drug delivery.

In addition to its biological relevance, the transport dynamics of polymers
through nanopores is of fundamental interest in physics and chemistry. There
exists a flurry of experimental~\cite{Akeson,Meller00,Meller01,Evilevitch} and
theoretical
~\cite{Sung,Park98,Muthukumar,Muthukumar01,Lubensky,Metzler,Ambj,Zandi,Chuang,Kantor,Grosberg,Luijten,
Gopinathan,Panja,Dubbeldam,Milchev,Luo1,Luo2,Huopaniemi,Luo3,Luo4,Santtu}
studies devoted to this subject. The passage of a flexible chain through a
nanopore involves a large entropic barrier, thus polymer translocation needs
driving forces, which can be provided by an external applied electric field in
the pore
~\cite{Kasianowicz,Akeson,Meller00,Meller01,Lubensky,Luo2,Luo3,Luo4,Santtu}, a
pulling force exerted on the end of a polymer~\cite{Santtu,Kantor,Huopaniemi},
binding particles (chaperones)~\cite{Ambj,Zandi}, or geometrical confinement of
the polymer~\cite{Evilevitch,Park98,Muthukumar01,Luijten,Panja,Gopinathan}.

\begin{figure}
  \includegraphics*[width=\figurewidth]{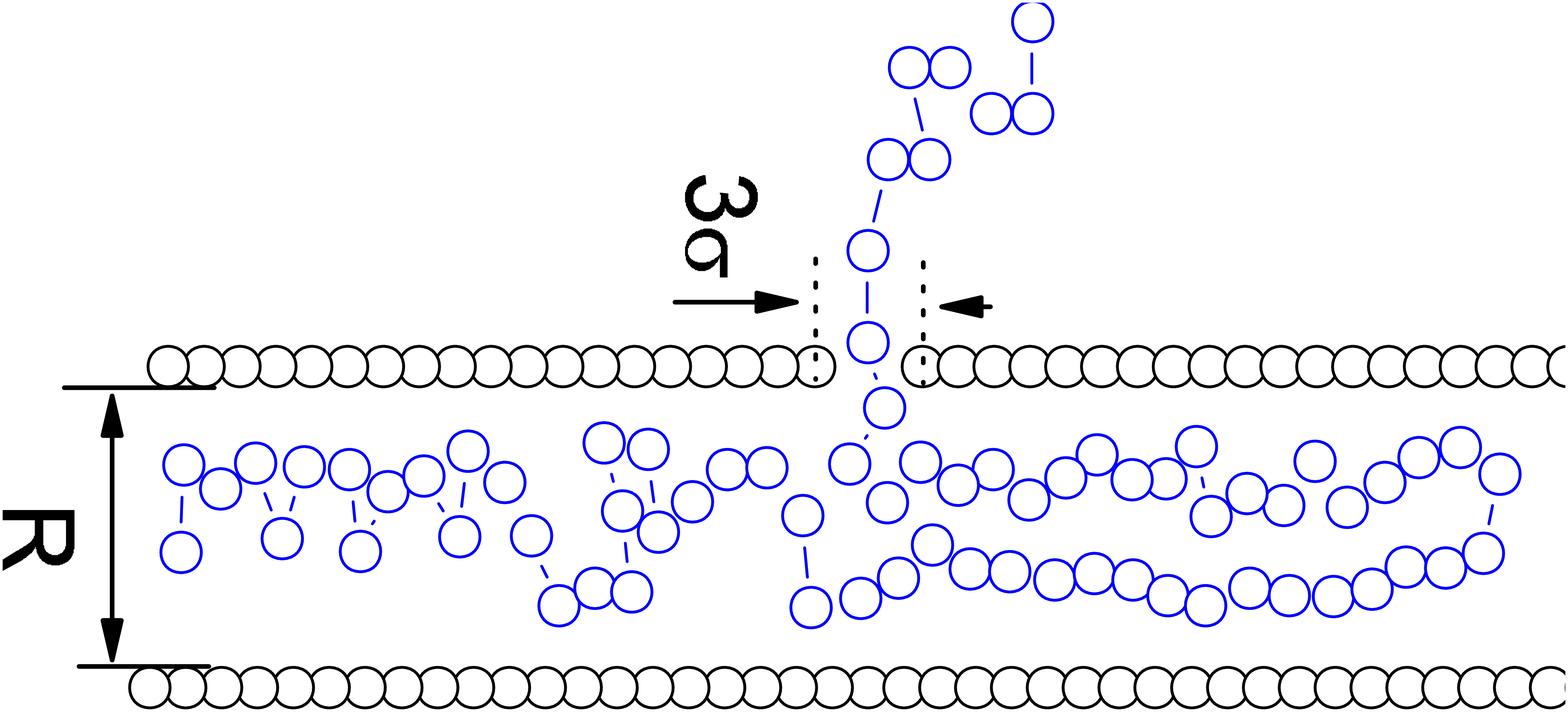}
\caption{A schematic representation of the system in 2D.
The simulations are carried out in both 2D and 3D.
For a planar confinement (3D), two plates are separated by a distance $R$.
One wall has a pore of length $L=\sigma$ and diameter $W=3\sigma$.
For polymers confined between two strips (2D) the pore width is $3\sigma$.
        }
 \label{Fig1}
\end{figure}

Among these mechanisms, less attention has been paid to confinement-driven
translocation. In particular its dynamics remains unclear, despite its
importance to biological processes including viral ejection, drug delivery,
controlled release from a nanochannel, etc. We here investigate the generic
behavior of polymer release through a small pore from a confined environment
(Fig.~\ref{Fig1}).

Based on Kantor and Kardar's scaling arguments for unhindered motion of the
chain~\cite{Kantor}, Cacciuto and Luijten~\cite{Luijten} suggested that $\tau
\sim N^{1+\nu}R^{1/\nu}$ for planar confinement shown in Fig. \ref{Fig1}, where
$\nu$ is the Flory exponent~\cite{de Gennes,Doi,Rubinstein} ($\nu=0.588$ in 3D)
and $R$ the separation between the walls. Panja \textit{et al}.~\cite{Panja}
considered translocation out of planar confinement as two-dimensional version
of translocation through a nanopore with an electric field applied in the pore,
and suggested $\tau \sim N^{2\nu_{2D}}$ ($\nu_{2D}=3/4$ in 2D), in contrast to
above scaling exponent $1+\nu$. To solve this contradiction, in this work we
investigate the dynamics of polymer translocation out of planar confinement
(3D) and out of two strips (2D) using both analytic scaling arguments and
Langevin dynamics simulations.
As in the above mentioned theories, we also ignore the hydrodynamic
interactions here. Our scaling arguments include geometric effects that have
been left out in previous studies, leading to a new set of scaling exponents of
translocation time $\tau$ with regard to polymer length $N$ as well as the size
of the confinement space $R$. The theoretical findings for $R$ dependence of
$\tau$ are fully supported by numerical simulation results both in 2D and 3D.
We consider a geometry as shown in Fig. \ref{Fig1}, where two walls with
separation $R$ are formed by stationary particles within a distance $\sigma$
from each other.

\section{Scaling argument}
%
A number of recent theories~\cite{Sung,Muthukumar,Chuang,Kantor} have been
developed for the dynamics of polymer translocation. Of them, Kantor and
Kardar~\cite{Kantor} provided lower bounds for both the translocation time for
pulling the polymer by the end and driving the polymer by a chemical potential
difference applied across the membrane. Essentially, the lower bound is the
time for the unimpeded polymer moving a distance of order of its size. For
driven translocation with a chemical potential difference applied across the
membrane, there is a force of $\Delta \mu /\sigma$ applied to the few monomers
inside the pore. As a consequence, Kantor and Kardar\cite{Kantor} argued that
its center of mass should move with a velocity $v \sim \Delta \mu/N$. Thus the
lower bound for the translocation time of an unhindered polymer is the time to
move through a distance $R_g$ (radius of gyration of the polymer in an
unconfined state), and scales as $\tau \sim R_g/v\sim N^{1+\nu}/(\Delta \mu)$.

Now, let us focus on the translocation out of confined environments. According
to the blob picture~\cite{de Gennes}, a chain confined between two parallel
plates with separation $\sigma \ll R \ll R_g$ will form a 2D self-avoiding walk
consisting of $n_b$ blobs of size $R$. Each blob contains
$g=(R/\sigma)^{1/\nu}$ monomers and the number of blobs is
$n_b=N/g=N(R/\sigma)^{-1/\nu}$. The free energy cost of the confined chain in
units of $k_BT$, $\Delta F$, is simply the number of the blobs. Thus the
chemical potential difference that provides a driving force for the
translocation process is given by
\begin{equation}
  \Delta \mu = \Delta F/N \sim R^{-1/\nu}.
\label{eq1}
\end{equation}
%
%
The remaining ingredients to complete the scaling argument are the
length scale $L_0$ through which the polymer has to move during the
entire translocation process and the average translocation velocity
$v$. As a result, the translocation time can be estimated as $\tau
\sim L_0/v$.
Cacciuto \textit{et al}.~\cite{Luijten} chose the radius of gyration
$R_g$ in the unconfined state for $L_0$ and $v \sim \Delta \mu/N$
and obtained the scaling result $\tau \sim N^{1+\nu}R^{1/\nu} \sim
N^{1.588}R^{1.70}$, while Panja \textit{et al}.~\cite{Panja}
criticized the scaling exponent $1+\nu$ for $N$ dependence of $\tau$
and considered the translocation out of the planar confinement as
the 2D version of translocation driven by a chemical potential
difference applied across the membrane. As a result, they obtained
the exponent $2 \nu_{2D}=1.50$ for the $N$ dependence in a 3D
system.

Here, we argue that the correct scaling results can only be obtained by
properly identifying the length scale $L_0$ and the scaling of the
translocation velocity $v$. Due to the highly non-equilibrium nature of the
driven translocation problem, the scaling of the average translocation velocity
$v$ with respect to the chain length $N$ is of the from $v \sim N^{\beta}$,
where the exponent $\beta \leq 1$~\cite{footnote}. For the present choice of
parameters, we find that $\beta \approx 0.8$ as will be demonstrated below.
%
%
For planar confinement (3D), the blob picture predicts the
longitudinal size of the polymer to be~\cite{Rubinstein,Doi,de
Gennes}
\begin{equation}
 R_{\parallel} \sim N^{\nu_{2D}} \sigma (\frac {\sigma} {R})^{\nu_{2D}/\nu-1}
                 \sim N^{3/4} \sigma (\frac {\sigma} {R})^{0.28},
\label{eq2}
\end{equation}
with $\nu_{2D}$ being the Flory exponent in 2D.
For successful translocation out of planar confinement, the polymer
has to move a distance $L_0 \sim R_{\parallel}$. Therefore, the
translocation time can be estimated as
\begin{equation}
 \tau  \sim  \frac {R_{\parallel}} {v}
      \sim N^{\beta+\nu_{2D}}R^{1+(1-\nu_{2D})/\nu}.
\label{eq3}
\end{equation}
Numerically, Eq. (\ref{eq3}) yields a scaling result of $\tau \sim
R^{1.43}$ for confinement driven translocation in 3D, which is
different from the prediction $\tau \sim R^{1.70}$ obtained by
Cacciuto \textit{et al}.~\cite{Luijten}. In fact, their numerical
results based on Monte Carlo simulations show $\tau \sim R^{1.54 \pm
0.10}$, which is close to our scaling prediction in Eq. (\ref
{eq3}).

Next, we also consider translocation out of confinement into a 2D
environment which has not been addressed
previously~\cite{Park98,Muthukumar01,Luijten,Panja}. For a polymer
confined between two strips embedded in 2D, the blob picture
predicts the longitudinal size of the chain to
be~\cite{Rubinstein,de Gennes}
\begin{equation}
 R_{\parallel} \sim N \sigma (\frac {\sigma} {R})^{-1+1/\nu_{2D}} \sim NR^{-1/3}.
\label{eq4}
\end{equation}
In this case the free energy excess in Eq.~(\ref{eq1}) is valid if
$\nu$ is replaced by $\nu_{2D}$. Thus, the translocation time scales
as
\begin{equation}
 \tau \sim  \frac {R_{\parallel}} {v} \sim \frac {NR^{1-1/\nu_{2D}}} {N^{-\beta}R^{-1/\nu_{2D}}}
      \sim N^{\beta+1}R 
      ,
\label{eq5}
\end{equation}
showing a linear dependence on $R$. Both predictions in
Eqs.~(\ref{eq3}) and (\ref{eq5}) for $R$ dependence are confirmed by
simulations below.

\section{Model and Methods}
In our numerical simulations, the polymer chains are modeled as
bead-spring chains of Lennard-Jones (LJ) particles with the Finite
Extension Nonlinear Elastic (FENE) potential. Excluded volume
interaction between monomers is modeled by a short range repulsive
LJ potential: $U_{LJ} (r)=4\epsilon
[{(\frac{\sigma}{r})}^{12}-{(\frac{\sigma} {r})}^6]+\epsilon$ for
$r\le 2^{1/6}\sigma$ and 0 for $r>2^{1/6}\sigma$. Here, $\sigma$ is
the diameter of a monomer, and $\epsilon$ is the potential depth.
The connectivity between neighboring monomers is modeled as a FENE
spring with $U_{FENE} (r)=-\frac{1}{2}kR_0^2\ln(1-r^2/R_0^2)$, where
$r$ is the distance between consecutive monomers, $k$ the spring
constant, and $R_0$ the maximum allowed separation between connected
monomers.
Between all monomer-wall particle pairs, there exists the same short
range repulsive LJ interaction as described above.

In the Langevin dynamics simulation, each monomer is subjected to
conservative, frictional, and random forces, respectively,
with~\cite{Allen} $m{\bf \ddot {r}}_i =-{\bf
\nabla}({U}_{LJ}+{U}_{FENE}) -\xi \dot{\bf r}_i + {\bf F}_i^R$,
where $m$ is the monomer's mass, $\xi$ is the friction coefficient
and ${\bf F}_i^R$ is the random force which satisfies the
fluctuation-dissipation theorem. In the present work, we use the LJ
parameters $\epsilon$ and $\sigma$ and the monomer mass $m$ to fix
the energy, length and mass scales respectively. Time scale is then
given by $t_{LJ}=(m\sigma^2/\epsilon)^{1/2}$. The dimensionless
parameters in our simulations are $R_0=2$, $k=7$, $\xi=0.7$ and
$k_{B}T=1.2$. The Langevin equation is integrated in time by a
method described by Ermak and Buckholz~\cite{Ermak} in both 3D and
2D.
To create the initial configuration, the first monomer of the chain
is placed in the entrance of the pore, while the remaining monomers
are initially squeezed into the space between two plates (3D) and
the space between two strips (2D) under thermal collisions described
by the Langevin thermostat to obtain an equilibrium configuration.
Typically, we average our data over 1000 independent runs.

According to the definition of the translocation time, at the
completion of the translocation process, the chain has moved a
distance of $R_{\parallel}$ along the direction perpendicular to the
axis of the pore. We now define the translocation velocity as $v=
\langle R_{\parallel} \rangle /\tau$. Using the definition $v=
\langle {R_{\parallel}}_i /\tau_i \rangle$ for the translocation
velocity, we observed similar results. Here ${R_{\parallel}}_i$ and
$\tau_i$ denote the values of $R_{\parallel}$ and $\tau$ for each
successful run.

\section{Numerical results}

\begin{figure}
\includegraphics*[width=\figurewidth]{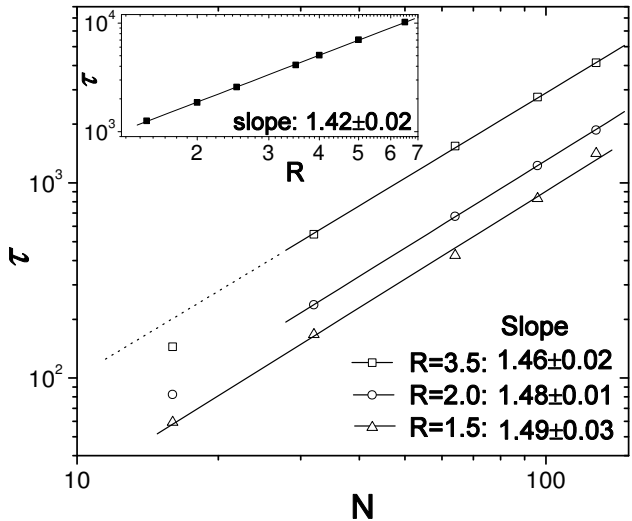}
\caption{Translocation time $\tau$ as a function of the chain length $N$ for
different $R$ in 3D. The insert shows $\tau$ as a function of $R$ for chain
length $N=128$ in 3D.
        }
\label{Fig2}
\end{figure}

For translocation out of a planar confinement, the results are shown in Fig.
\ref{Fig2}. The insert of Fig. \ref{Fig2} shows the $R$ dependence of the
translocation time. The scaling exponent is $1.42\pm0.02$, which is in good
agreement with the exponent 1.43 from our scaling prediction in Eq. (\ref
{eq3}).
For $\tau$ with $N$ for $R=$ 3.5, 2.0 and 1.5, we get the scaling exponents of
$1.46\pm0.02$, $1.48\pm0.01$, and $1.49\pm0.03$, respectively. With decreasing
$R$, the exponent slightly increases.
As to translocation velocity, such as for $R=1.5$ we get $\beta=0.77\pm0.01$.
According to Eq. (\ref {eq3}), $\beta+\nu_{2D}=1.52$, which is very close to
1.49. These results demonstrate that the scaling arguments for unimpeded
translocation provide an accurate estimate for the translocation time. Although
the scaling exponent for $N$ dependence is quite close to the value 1.55
obtained by Cacciuto \textit{et al}.~\cite{Luijten}, the physical origin is
quite different.

\begin{figure}
\includegraphics*[width=\figurewidth]{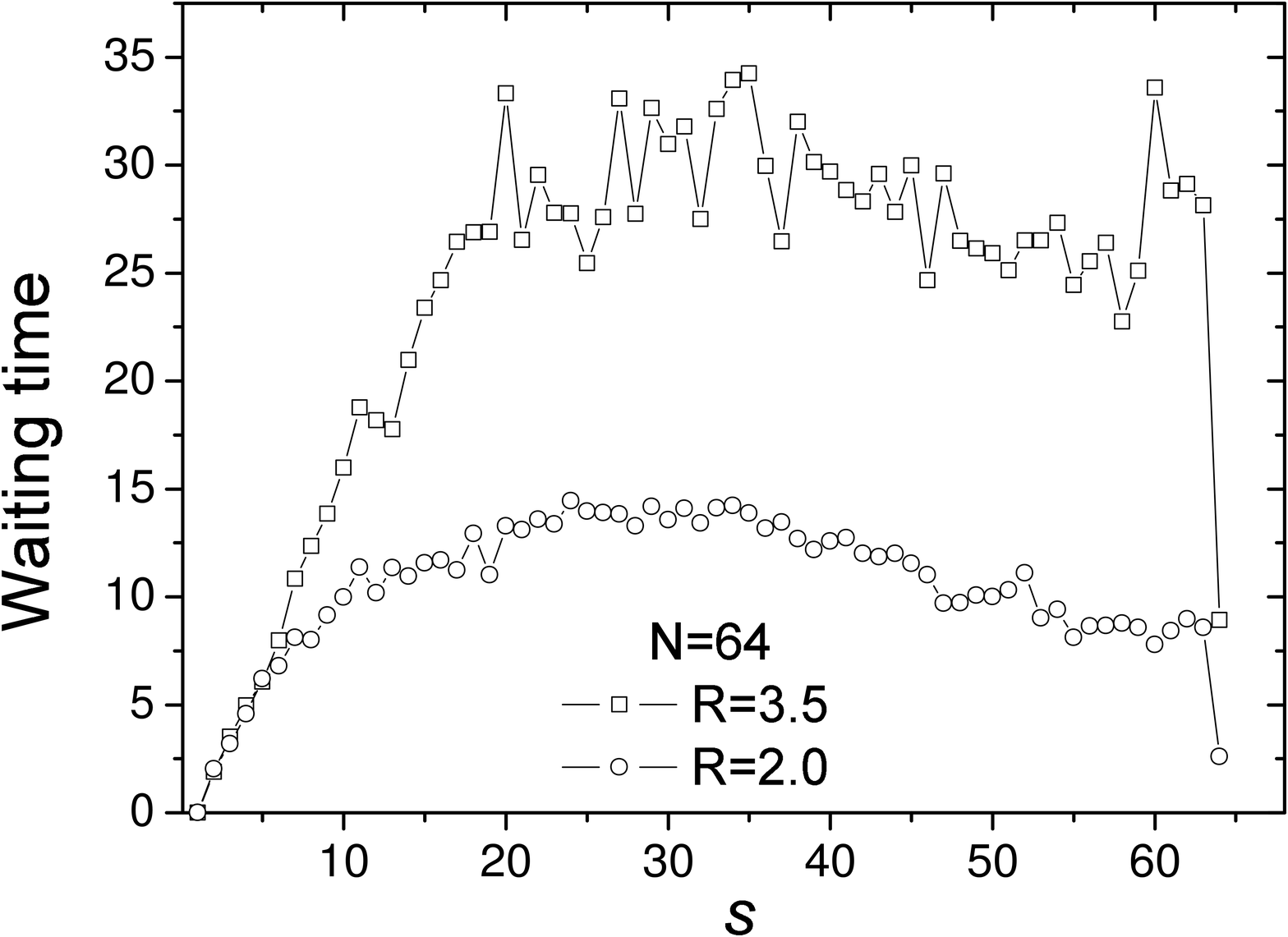}
\caption{Waiting time distribution for 3D simulations.
        }
\label{Fig3}
\end{figure}

The dynamics of a single segment passing through the pore during translocation
is an important quantity considerably affected by different driving mechanisms.
The nonequilibrium nature of translocation has a significant effect on it. We
have numerically calculated the waiting times for all monomers in a chain of
length $N$. We define the waiting time of monomer $s$ as the average time
between the events that monomer $s$ and monomer $s+1$ exit the pore.
In our previous work~\cite{Luo2,Huopaniemi} studying translocation driven by a
voltage across the pore without confinement on either side, we found that for
short polymers, such as $N=100$, the monomers in the middle of the polymer need
the longest time to translocate and the distribution is close to symmetric with
respect to the middle monomer~\cite{Luo2}. Fig. \ref{Fig3} shows the waiting
time distributions for translocation out of planar confinements with $R=2.0$
and 3.5 for $N=64$. Compared with the unconfined potential driven case, the
waiting time distribution is quite different. The waiting times increase until
$s \sim 20$, and then almost saturate. This can be qualitatively understood as
follows. The initial rise of the waiting time has the same origin as the
potential driven case, with $\Delta \mu$ given by Eq. (\ref {eq1}) playing the
role of the applied voltage. Here, the driving force has two components, one
from the cost of the free energy due to blobs and the second due to entropic
wall repulsion for the translocated beads. These are balanced by frictional
force in the pore. The reduction in the overall driving causes the initial rise
of the waiting time distribution, which is eventually balanced by decreasing
frictional force leading to a plateau in the waiting time distribution.

\begin{figure}
\includegraphics*[width=\figurewidth]{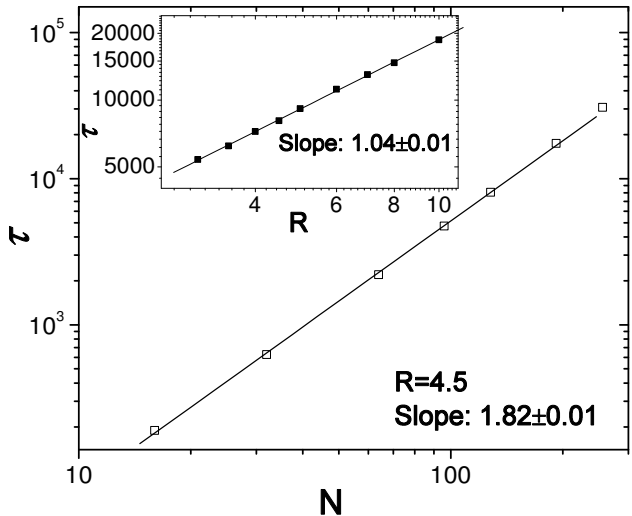}
\caption{Translocation time $\tau$ as a function of the chain length $N$ for $R=4.5$ in 2D.
The insert shows $\tau$ as a function of $R$ for chain length $N=128$ in 2D.
        }
\label{Fig4}
\end{figure}

For translocation out of two strips, the results are shown in Fig. \ref{Fig4}.
The translocation time increases linearly with increasing $R$ with the scaling
exponent $1.04 \pm 0.01$, see the insert of Fig. \ref{Fig4}, which is in
excellent agreement with the prediction in Eq. (\ref {eq5}). Moreover, for
$\tau$ as a function of $N$, we get scaling exponent of $1.82\pm0.01$ for
$R=4.5$.
For translocation velocity, we get $\beta=0.79 \pm 0.02$. According to Eq.
(\ref {eq5}), $\beta+1=1.79$, which is very close to the scaling exponent
$1.82$.

\begin{figure}
\includegraphics*[width=\figurewidth]{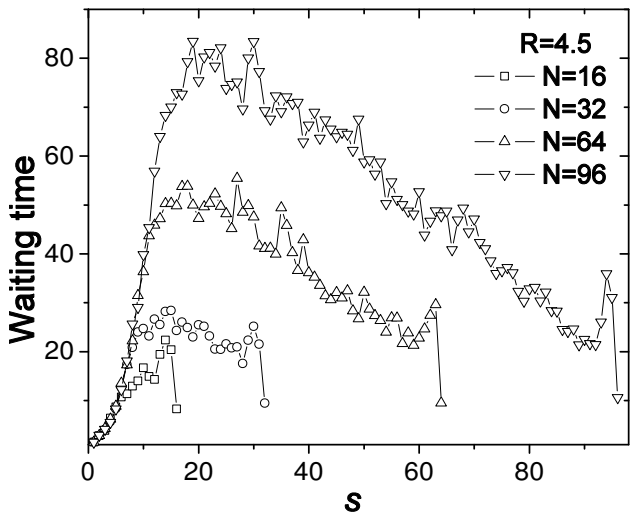}
\caption{Waiting time distribution for 2D simulations.
        }
\label{Fig5}
\end{figure}

Fig. \ref{Fig5} shows the waiting time distributions for translocation out of
two strips with $R=4.5$ for $N=16$, $32$, $64$ and $96$. For $N \ge 64$, the
waiting time increases with $s$ until the maximum at $s \sim 20$, and then
almost linearly decreases with $s$. For all chain lengths, the monomers at the
end of the chain from $s=N-4$ to $N-1$ still need a little longer time to
translocate due to very weak confinement. The observed behavior is due to the
interplay of decreasing $\Delta \mu$ and increasing pulling entropic force. The
balance of these two factors is differs from 2D to 3D.

\section{Conclusions} \label{chap-conclusions}

In this work, we have considered the dynamics of polymer translocation out of
confined environments based on both scaling arguments and Langevin dynamics
simulations. Analytic scaling arguments lead to the prediction that the
translocation time scales like $\tau\sim
N^{\beta+\nu_{2D}}R^{1+(1-\nu_{2D})/\nu}$ for translocation out of a planar
confinement between two walls with separation $R$ into a 3D environment, and
$\tau \sim N^{\beta+1}R$ for translocation out of two strips with separation
$R$ into a 2D environment. Here, $N$ is the chain length, $\nu$ and $\nu_{2D}$
are the Flory exponents in 3D and 2D, and $\beta$ is the scaling exponent of
translocation velocity with $N$, whose value for the present choice of
parameters is $\beta \approx 0.8$ based on Langevin dynamics simulations. These
scaling exponents improve on the previously reported
results~\cite{Luijten,Panja}.

\begin{acknowledgments}

This work has been supported in part by the Deutsche Forschungsgemeinschaft
(DFG). T.A.N. acknowledges support from The Academy of Finland through its
Center of Excellence (COMP) and TransPoly Consortium grants. We also
acknowledge CSC Ltd. for allocation of computational resources.

\end{acknowledgments}


\begin{thebibliography}{8}

\bibitem{Alberts} B. Alberts and D. Bray, J. Lewis, M. Raff, and J. D. Watson,
                              \textit{Molecular Biology of the Cell}(Garland, New York, 1994).
\bibitem{Kasianowicz} J. J. Kasianowicz, E. Brandin, D. Branton and D. W. Deamer,
                                \textit{Proc. Natl. Acad. Sci. U.S.A.} {\bf 93}, 13770 (1996).
\bibitem{Meller03} A. Meller, \textit{J. Phys.: Condens. Matter} {\bf 15}, R581 (2003).
\bibitem{Akeson} M. Akeson, D. Branton, J. J. Kasianowicz, E. Brandin, and D. W. Deamer,
                                             \textit{Biophys. J.} {\bf 77}, 3227 (1999).
\bibitem{Meller00} A. Meller, L. Nivon, E. Brandin, J. A. Golovchenko, and D. Branton,
                             \textit{Proc. Natl. Acad. Sci. U.S.A.} {\bf 97}, 1079 (2000).
\bibitem{Meller01} A. Meller, L. Nivon, and D. Branton, \textit{Phys. Rev. Lett.} {\bf 86}, 3435 (2001).
\bibitem{Evilevitch} A. Evilevitch, L. Lavelle, C. M. Knobler, E. Raspaud, and W. M. Gelbart,
                                \textit{Proc. Natl. Acad. Sci. U.S.A.} {\bf 100}, 9292 (2003).
\bibitem{Sung}   W. Sung and P. J. Park, \textit{Phys. Rev. Lett.} {\bf 77}, 783 (1996).
\bibitem{Park98} P. J. Park and W. Sung, \textit{Phys. Rev. E}  {\bf 57}, 730 (1998).
\bibitem{Muthukumar}   M. Muthukumar, \textit{J. Chem. Phys.} {\bf 111}, 10371 (1999)
\bibitem{Muthukumar01}   M. Muthukumar, \textit{Phys. Rev. Lett.} {\bf 86}, 3188 (2001).
\bibitem{Lubensky} D. K. Lubensky and D. R. Nelson, \textit{Biophys. J.} {\bf 77}, 1824 (1999).
%
\bibitem{Metzler} R. Metzler and J. Klafter, \textit{Biophys. J.} {\bf 85}, 2776 (2003).
\bibitem{Ambj} T. Ambj{\"o}rnsson and R. Metzler, \textit{Phys. Biol.} {\bf 1}, 19 (2004);
               T. Ambj{\"o}rnsson, M. A. Lomholt, and R. Metzler,
                                  \textit{J. Phys.: Condens. Matter} {\bf 17}, S3945 (2005).
\bibitem{Zandi} R. Zandi, D. Reguera, J. Rudnick, and W. M. Gelbart,
                                    \textit{Proc. Natl. Acad. Sci. U.S.A.} {\bf 100}, 8649 (2003).
\bibitem{Chuang} J. Chuang, Y. Kantor, and M. Kardar, \textit{Phys. Rev. E} {\bf 65}, 011802 (2002);
                 Y. Kantor and M. Kardar, \textit{Phys. Rev. E} {\bf 76}, 061121 (2007);
                 C. Chatelain, Y. Kantor and M. Kardar, \textit{Phys. Rev. E} {\bf 78}, 021129 (2008).
\bibitem{Kantor} Y. Kantor and M. Kardar,  \textit{Phys. Rev. E} {\bf 69}, 021806 (2004).
\bibitem{Grosberg} A. Yu. Grosberg, S. Nechaev, M. Tamm, O. Vasilyev,
                                               \textit{Phys. Rev. Lett.} {\bf 96}, 228105 (2006).
\bibitem{Luijten} A. Cacciuto and E. Luijten,  \textit{Phys. Rev. Lett.} {\bf 96}, 238104 (2006).
\bibitem{Panja} D. Panja, G. T. Barkema, and R. C. Ball,
                           \textit{J. Phys.: Condens. Matter} {\bf 20}, 075101 (2008).
\bibitem{Gopinathan} A. Gopinathan and Y. W. Kim, \textit{Phys. Rev. Lett.} {\bf 99}, 228106 (2007).
\bibitem{Dubbeldam} J. L. A. Dubbeldam, A. Milchev, V.G. Rostiashvili, and T.A. Vilgis,
                                    \textit{Phys. Rev. E} {\bf 76}, 010801(R) (2007);
                                    \textit{Europhys. Lett.} {\bf 79}, 18002 (2007).
\bibitem{Milchev} A. Milchev, K. Binder, and A. Bhattacharya, \textit{J. Chem. Phys.} {\bf 121}, 6042 (2004).
%
\bibitem{Luo1} K. F. Luo, T. Ala-Nissila, and S. C. Ying, \textit{J. Chem. Phys.} {\bf 124}, 034714 (2006).
\bibitem{Huopaniemi} I. Huopaniemi, K. F. Luo, T. Ala-Nissila, and S. C. Ying,
                                          \textit{Phys. Rev. E} {\bf 75}, 061912 (2007).
\bibitem{Luo2} K. F. Luo, I. Huopaniemi, T. Ala-Nissila, and S. C. Ying,
                                   \textit{J. Chem. Phys.} {\bf 124}, 114704 (2006);
               I. Huopaniemi, K. F. Luo, T. Ala-Nissila, and S. C. Ying,
                                   \textit{ibid} {\bf 125}, 124901 (2006).
\bibitem{Luo3} K. F. Luo, T. Ala-Nissila, S. C. Ying, and A. Bhattacharya,
              \textit{J. Chem. Phys.} {\bf 126}, 145101 (2007); \textit{Phys. Rev. Lett.} {\bf 99}, 148102 (2007);
              \textit{ibid} {\bf 100}, 058101 (2008); \textit{Phys. Rev. E} {\bf 78}, 061911 (2008);
              \textit{ibid} {\bf 78}, 061918 (2008).
\bibitem{Luo4} K. F. Luo, S. T. T. Ollila, I. Huopaniemi, T. Ala-Nissila, P. Pomorski, M. Karttunen, S. C. Ying,
               and A. Bhattacharya, \textit{Phys. Rev. E} {\bf 78}, 050901(R) (2008).
\bibitem{Santtu} S. T. T. Ollila, K. F. Luo, T. Ala-Nissila, and S. C. Ying,
                \textit{Eur. Phys. J. E} {\bf 28}, 385 (2009).
\bibitem{footnote} Recent numerical results show that for translocation with driving force localized
in the pore, the translocation velocity $v$ scales as $1/N^{\beta}$ with $\beta \approx 0.8$ in 3D
instead of $1/N$ for highly non-equlibrium translocation processes (A. Bhattacharya \textit{et
al.}, arXiv:0808.1868; K. Luo \textit{et al.}, to be published).
\bibitem{de Gennes} P. G. de Gennes, \textit{Scaling Concepts in Polymer Physics}
                                               (Cornell University Press, Ithaca, NY, 1979);
  M. Daoud, and P. G. de Gennes, \textit{J. Physique} {\bf 38}, 85 (1977).
\bibitem{Doi} M. Doi, and S. F. Edwards, \textit{The Theory of Polymer Dynamics} (Clarendon, Oxford, 1986).
\bibitem{Rubinstein} M. Rubinstein, and R. Colby, \textit{Polymer Physics} (Oxford University Press, Oxford, 2003).
%
\bibitem{Allen} M.P. Allen, D.J. Tildesley, \textit{Computer Simulation of Liquids} (Oxford University Press, 1987).
\bibitem{Ermak} D. L. Ermak and H. Buckholz, \textit{J. Comput. Phys.} {\bf 35}, 169 (1980).
\end{thebibliography}
\end{document}